\documentclass[twoside]{dis07}
\usepackage[latin1]{inputenc}
\usepackage[dvips]{graphicx,epsfig,color}
\usepackage{wrapfig,rotating}
\usepackage{amssymb,amsmath,array}

\begin{document}
\title{Novel Master Formula for Twist-3 Soft-Gluon-Pole Mechanism 
to Single Transverse-Spin Asymmetry}

\author{Yuji Koike$^1$ and Kazuhiro Tanaka$^2$
%
\thanks{Supported by the Grant-in-Aid for Scientific Research No. B-19340063.}
%
\vspace{.3cm}\\
%
1- Department of Physics, Niigata University, Ikarashi, Niigata 950-2181, Japan
%
\vspace{.1cm}\\
2- Department of Physics, Juntendo University, Inba-gun, Chiba 270-1695, Japan}

\maketitle

\begin{abstract}
We prove that twist-3 soft-gluon-pole (SGP) cross section for single
spin asymmetries 
is determined by a certain ``primordial'' twist-2 cross section
up to kinematic and 
color factors in the leading order perturbative QCD. 
This unveils universal structure behind the SGP cross sections
in a variety of hard processes,
and also the special role
of the scale invariance in the corresponding primordial cross section,
which leads to remarkable simplification of the SGP cross sections 
for the production of massless particle, such as those for
pion production 
$p^\uparrow p\to \pi X$ and
direct-photon production 
$p^\uparrow p\to \gamma X$.
\end{abstract}

The single transverse-spin asymmetry (SSA)
is observed as ``T-odd'' effect proportional 
to $\vec{S}_\perp \cdot(\vec{p} \times \vec{q})$
in the cross section for the scattering of transversely polarized proton with momentum $p$
and spin $S_\perp$,
off unpolarized particle with momentum $p'$, producing a particle with momentum $q$ 
which is observed in the final state.
Famous examples~\cite{HC} are $p^\uparrow p\to\pi X$ with the large asymmetry $A_N \sim 0.3$
observed in the forward direction, 
and semi-inclusive deep 
inelastic scattering (SIDIS), $ep^\uparrow\to e\pi X$, by HERMES and COMPASS Collaborations.
The Drell-Yan (DY) process, $p^\uparrow p\to \ell^+ \ell^- X$,
and the direct $\gamma$ production, $p^\uparrow p\to \gamma X$,
at RHIC, J-PARC, etc. are also expected to play important roles
for the study of SSA.

The SSA requires,
(i) nonzero $q_\perp$ originating 
from transverse motion
of quark or gluon; 
(ii) proton helicity flip; 
and (iii) interaction beyond Born level to produce the
interfering phase for the cross section. 
For processes with $q_\perp\sim \Lambda_{\rm QCD}$, all (i)-(iii) may be generated 
nonperturbatively from 
the T-odd, transverse-momentum-dependent parton distribution/fragmentation 
functions~\cite{boer}.
For large $q_\perp \gg \Lambda_{\rm QCD}$,
(i) should come from perturbative mechanism, while the nonperturbative effects 
can participate in the other two, (ii) and (iii), allowing us to obtain large SSA.
This is realized as the twist-3 mechanism in QCD for the SSA.
Recently we have thoroughly discussed the collinear-factorization property and gauge invariance
in the twist-3 mechanism in the context of the SSA in SIDIS~\cite{ekt06}.
We have also revealed universal structure behind the twist-3 mechanism~\cite{KT,KT2}, 
which we discuss here.

\begin{wrapfigure}{r}{0.5\columnwidth}
\centerline{\includegraphics[height=.12\textheight,clip]{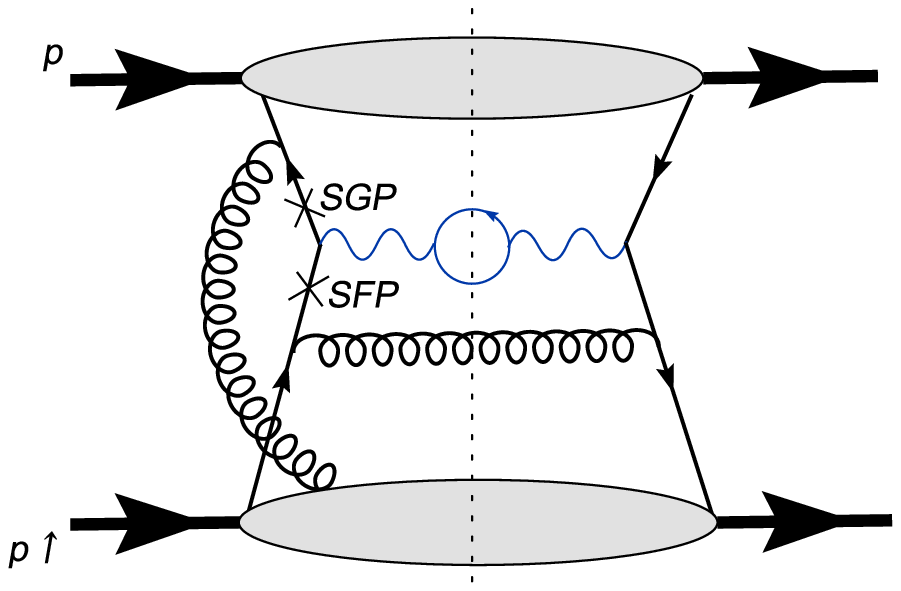}\hspace{0.1cm}
\includegraphics[height=.12\textheight,clip]{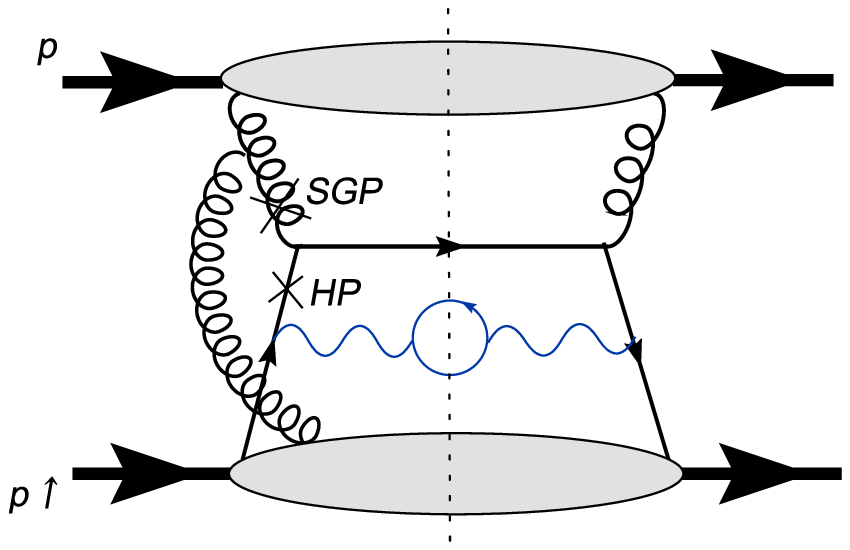}}
\caption{
Typical 
diagrams for DY 
process 
with $q_\perp \gg \Lambda_{\rm QCD}$.
The cross $\times$ denotes 
the pole contribution of the parton propagator.
}
\end{wrapfigure}
As an example, consider 
the DY production of the dilepton with $q_\perp \gg \Lambda_{\rm QCD}$:
The large $q_\perp$ of (i) is provided by hard interaction
as the recoil from a hard final-state parton, as illustrated in Fig.~1.
Proton helicity flip (ii) is provided by the participation of the coherent, nonperturbative gluon
from the transversely polarized proton, the lower blob in Fig.~1, associated with
the twist-3 quark-gluon correlation functions such as $G_{F}(x_1, x_2)$ 
with $x_1$ ($x_2$)
the lightcone momentum 
fraction of the quark leaving from (entering into) the proton~\cite{ekt06}.
Due to the coupling of this coherent gluon,
some parton propagators in the partonic subprocess can be on-shell,
and this produces the imaginary phase of (iii) as the pole contribution using
$1/(k^2 + i\varepsilon) = P(1/k^2) - i\pi\delta(k^2)$.
Depending on the value of the coherent-gluon's momentum $k_g$ at the corresponding poles, 
these 
are soft-gluon pole (SGP) for $k_g=0$, and soft-fermion pole (SFP)
and hard pole (HP) for $k_g \neq 0$.

Among these three-types of poles, the SGP deserves special attention; indeed,
the SGP is considered to give dominant twist-3 mechanism in many phenomenological applications
(see e.g. \cite{ekt06,KT}). 
We have derived the {\em master formula} for the SGP cross section, which embodies the remarkable structure 
that the SGP contributions from many Feynman diagrams of the type of Fig.~1
are united into a derivative of the 2-to-2 partonic Born diagrams without the coherent-gluon line: 
The SSA for the DY process can be expressed as~\cite{KT} 
\begin{equation}
\frac{d\sigma^{\rm DY}_{\mbox{\scriptsize tw-3,SGP}}}{[d\omega]}= 
\frac{\pi M_N}{2C_F}
\epsilon^{\sigma pnS_\perp} \sum_{j=\bar{q},g}{\cal B}_j \int \frac{dx'}{x'}\int \frac{dx}{x} f_j(x') 
\frac{\partial H_{jq}(x', x)}{\partial (x' p^{\prime \sigma})} G_F^q(x,x),
\label{tw3formula}
\end{equation}
where $j= \bar{q}$ and $g$ represent the ``$q\bar{q}$-annihilation'' 
and ``$qg$-scattering'' channels, respectively,
corresponding to the left and right diagrams in Fig.~1. 
$f_{j}(x')$ denotes the twist-2 parton distribution functions for the unpolarized proton, and
$M_N$ is the nucleon mass representing nonperturbative scale
associated with the twist-3 correlation function $G_F^q(x,x)$ for the flavor $q$.
The sum over quark and antiquark flavors
is implicit for 
the index $q$ as $q=u, \bar{u}, d, \bar{d}, \cdots$.
$[d\omega]=dQ^2 dy d^2 q_\perp$ denotes the relevant differential elements 
with $Q^2 = q^2$ and $y$ the rapidity of the virtual photon.
The color factors are introduced as 
${\cal B}_q = 1/( 2N_c )$ and $-1/(2N_c )$ for quark and antiquark flavors, 
respectively,
${\cal B}_g = N_c/2$, and $C_F = (N_c^2-1)/(2N_c)$.
The derivative with respect to $p^{\prime}$ 
is taken under $p^{\prime 2}=0$, and
$H_{jq}(x', x)$ 
denote the partonic hard-scattering functions 
which participate in the unpolarized twist-2 cross section for 
DY process as
\begin{equation}
\frac{d\sigma^{\rm unpol,DY}_{\mbox{\scriptsize tw-2}}}{[d\omega]}= 
\sum_{j=\bar{q},g}  \int \frac{dx'}{x'} \int \frac{dx}{x} 
f_j(x') H_{jq}(x', x) f_q(x).
\label{tw2formula}
\end{equation}
Namely, in order to obtain the explicit formula for the twist-3 SGP contributions to the SSA,
knowledge of the twist-2 unpolarized cross section is sufficient.

A proof of (\ref{tw3formula}) is described in detail in \cite{KT}:
Only the {\em initial-state interaction} (ISI) diagrams like Fig.~1,
where
the coherent gluon couples
to an ``external parton'' associated with an initial-state hadron,
survive as the SGP contributions for DY process,
while the SGPs from the other diagrams cancel out combined with 
the corresponding ``mirror'' diagrams~\cite{JQVY06}. 
For such ISI diagrams,
the coherent-gluon insertion and the associated SGP structure
can be disentangled from the partonic subprocess,
keeping the remaining factors mostly intact.
For the {\em scalar-polarized} coherent-gluon,
this is performed using Ward identity; moreover,
also for the transversely-polarized coherent-gluons that 
are relevant at the twist-3 level, 
this can be performed exactly 
through the logic different from the scalar-polarized case~\cite{KT}.
Using the background field gauge,
the three-gluon coupling relevant to the $qg$-scattering channel can be disentangled similarly
as the quark-gluon coupling case.
After disentangling ISI with the coherent gluons, 
the collinear expansion 
in terms of the parton transverse momenta gives
the twist-3 cross section (\ref{tw3formula}) at the SGP,
as the response of 2-to-2 partonic 
subprocess
to the change of the transverse momentum 
carried by the external parton, 
to which the coherent gluon had coupled.
Note that ${\cal B}_{\bar{q}}$ and ${\cal B}_g$
in (\ref{tw3formula}) 
come from the insertion of the color matrix $t^a$ in the fundamental and adjoint 
representations, respectively, into the 2-to-2 subprocess, as the remnant of the coherent-gluon insertion
via the quark-gluon and three-gluon vertices.

The hard-scattering functions 
in the twist-2 factorization formula (\ref{tw2formula}) are expressed as
$H_{jq} (x', x)= ( \alpha_{em}^2 \alpha_s e_q^2 / 3\pi N_c s Q^2 )
\widehat{\sigma}_{jq}(\hat{s}, \hat{t}, \hat{u}) 
\delta \left( \hat{s}+\hat{t}+\hat{u}-Q^2 \right)$,
where $s=(p+p')^2$, and explicit form of $\widehat{\sigma}_{jq}(\hat{s}, \hat{t}, \hat{u})$ 
in terms of 
the partonic Mandelstam variables $\hat{s}, \hat{t}$ and $\hat{u}$
can be found in Eq.~(28) in \cite{KT}. The derivative in (\ref{tw3formula}) 
can be performed through that for the $\hat{u}$, and this may act on either 
$\widehat{\sigma}_{jq}$
or the delta function in 
$H_{jq}(x', x)$, where the latter case produces 
the derivative of the twist-3 correlation 
function, 
$dG_F^q (x, x)/dx$, as well as the non-derivative term $\propto G_F^q (x, x)$,
by partial integration with respect to $x$. 
Our master formula (\ref{tw3formula}) yields~\cite{KT}
\begin{eqnarray}
\lefteqn{
\frac{d\sigma^{\rm DY}_{\mbox{\scriptsize tw-3,SGP}}}{dQ^2 dy d^2 q_\perp}= 
\frac{\alpha_{em}^2 \alpha_s e_q^2}{3\pi N_c s Q^2}
\frac{\pi  M_N}{C_F} 
\epsilon^{pnS_\perp q_\perp}
\sum_{j=\bar{q}, g}{\cal B}_{j}
\int \frac{dx'}{x'} \int \frac{dx}{x} \delta \left( \hat{s}+\hat{t}+\hat{u}-Q^2 \right)
f_j(x') 
}
\nonumber\\
&&\;\;\;\;\;\;\times
\left\{  \frac{\widehat{\sigma}_{jq}}{-\hat{u}}  x\frac{dG_F^q (x, x)}{dx}
+\left[\frac{\widehat{\sigma}_{jq}}{\hat{u}}  
-\frac{\partial \widehat{\sigma}_{jq}}{\partial \hat{u}}  
- \frac{\hat{s}}{\hat{u}}\frac{\partial  \widehat{\sigma}_{jq}}{\partial \hat{s}}  
-\frac{\hat{t}-Q^2}{\hat{u}}\frac{\partial  \widehat{\sigma}_{jq}}{\partial \hat{t}}  
\right] G_F^q (x, x)\right\}.
\label{dyf}
\end{eqnarray}
This novel expression not only reproduces 
the known pattern~\cite{JQVY06} that the partonic hard scattering functions associated 
with the derivative term
are directly proportional to those participating in the twist-2 unpolarized process,
$\widehat{\sigma}_{jq}$, but also
reveals the structure
that was hidden in the corresponding formula in the literature~\cite{JQVY06}:
the partonic hard-scattering functions associated 
with the non-derivative term are also
completely determined by $\widehat{\sigma}_{jq}$.

We obtain the SSA for the direct $\gamma$ production 
in the real-photon limit, $Q^2 \rightarrow 0$; here
only the massless particles participate 
in the 2-to-2 Born subprocess, so that the corresponding partonic cross sections
$\widehat{\sigma}_{jq}$ are
scale-invariant as 
$(\hat{u}\partial / \partial \hat{u}  
+\hat{s} \partial / \partial \hat{s}  
+ \hat{t} \partial  / \partial \hat{t} ) \widehat{\sigma}_{jq}=0$. 
Consequently, (\ref{dyf}) reduces to the compact structure 
where $-\widehat{\sigma}_{jq} /\hat{u}$  
appears
both for derivative and
non-derivative terms, 
as the coefficient for
the combination, $x dG_F^q (x, x)/dx - G_F^q (x, x)$~\cite{KT}.

The DY process
can be formally transformed into SIDIS, $ep^\uparrow\to e\pi X$,
{\em crossing}
the initial unpolarized proton
into the final-state pion with momentum $P_h$,
and the virtual photon into the initial-state one.
The proof of (\ref{tw3formula})
discussed above 
is unaffected by the
analytic continuation corresponding to this ``crossing transformation'':
$p' \rightarrow - P_{h}$, $x' \rightarrow 1/z$, $f_{\bar{q}}(x')\rightarrow D_q (z)$, 
$f_g(x')\rightarrow D_g (z)$, 
and $q^\mu \rightarrow -q^{\mu}$,
where $D_j(z)$ denote 
the twist-2 parton fragmentation functions for the final-state pion,
and the new $q^{\mu}$ gives $Q^2 =- q^2$.
Our master formula (\ref{tw3formula})
now gives
the SGP contribution to the SSA in SIDIS, which is actually associated with 
the corresponding {\em final-state interaction} (FSI) diagrams, as
(${\cal C}_{q} \equiv {\cal B}_{\bar{q}}, {\cal C}_g \equiv {\cal B}_g$)~\cite{KT}
\begin{equation}
\frac{d\sigma^{\rm SIDIS}_{\mbox{\scriptsize tw-3,SGP}}}{[d\omega]}= 
\frac{\pi M_N}{C_F z_f^2}
\epsilon^{pn S_\perp P_{h \perp}} \left. \frac{\partial}{\partial q_T^2}
\frac{d\sigma^{\rm unpol,SIDIS}_{\mbox{\scriptsize tw-2}}}{[d\omega]}
\right|_{f_q(x)\rightarrow G_F^q(x,x),\ D_j(z) \rightarrow {\cal C}_j zD_j (z)},
\label{sidis}
\end{equation}
in a frame where the 3-momenta $\vec{q}$ and $\vec{p}$ of the
virtual photon and the transversely polarized nucleon are collinear along the $z$ axis.
$[d\omega]= dx_{bj}dQ^2 dz_f dq_T^2 d\phi$,
where, as usual, $x_{bj}={Q^2/ (2p\cdot q)}$, $z_f={p\cdot P_h / p\cdot q }$,
$q_T = P_{h\perp}/z_f$,
and $\phi$ is the azimuthal angle between the lepton and hadron planes. 
The twist-2 unpolarized cross section in the RHS of (\ref{sidis}) is known to have
several terms with different $\phi$-dependence, 
proportional to $1, \cos \phi$, and $\cos 2\phi$, respectively (see \cite{ekt06}).
Performing the derivative in (\ref{sidis}) explicitly, the result
obeys the similar pattern as (\ref{dyf}) with derivative and non-derivative terms,
for each azimuthal dependence, and
unveils the structure 
behind the complicated formula 
obtained by direct evaluation of the diagrams~\cite{ekt06}.

Our master formula
can be extended to ``QCD-induced'' $pp$ collisions, like $p^\uparrow p\to \pi X$~\cite{KT2}. 
For example, the $qq \rightarrow qq$ scattering
subprocess induce
the 
twist-2 unpolarized cross section,
\begin{equation}
E_h{d^3\sigma^{pp \rightarrow \pi X}_{\mbox{\scriptsize tw-2}}\over d^3 P_h}
={\alpha_s^2 \over s}
\int{dz\over z^2}
{dx'\over x'}
{dx\over x}
\delta(\hat{s}+\hat{t}+\hat{u})
D_q(z) f_q(x') f_q(x) \widehat{\sigma}_{U} ( \hat{s},\hat{t},\hat{u}) ,
\label{twist2}
\end{equation}
for 
$p(p)+p(p')\to\pi(P_h)+ X$, 
where
$E_h\equiv P_h^0$, and 
$\widehat{\sigma}_{U}( \hat{s},\hat{t},\hat{u})
= (C_F /N_c ) (\hat{s}^2 +\hat{u}^2)/\hat{t}^2 +(C_F /N_c )
(\hat{s}^2 +\hat{t}^2)/\hat{u}^2 +(-2C_F /N_c^2 )\hat{s}^2/(\hat{t}\hat{u})$ 
is the $qq \rightarrow qq$ unpolarized cross section~\cite{KQVY06}. 
The 
SGP contribution from the interference between $qqg\to qq$ and $qq\to qq$ 
is generated by FSI and
ISI with the coherent gluon as in Fig.~2 (a) and (b), which can be treated
similarly as the SIDIS and DY cases, respectively, 
and yields~\cite{KT2} the twist-3 cross section for 
$p^\uparrow p\to \pi X$:
\begin{eqnarray}
E_h{d^3\sigma^{pp \rightarrow \pi X}_{\mbox{\scriptsize tw-3,SGP}}
\over d^3 P_h}&& \!\!\!\!\!\!\!\!\!\!\!
={\pi M_N\alpha_s^2 \over s} 
\int{dz\over z^2}
{dx'\over x'}
{dx\over x}\delta(\hat{s}+\hat{t}+\hat{u}) D_q(z) f_q(x')
\left[ x{d G_F^q(x,x) \over dx}-G_F^q(x,x)\right]
\nonumber\\
&&\!\!\!\!\!\!\!\!\times 
\left[ \frac{1}{z}\epsilon^{S_\perp P_h pn} +
{x' \hat{t}\over\hat{s}} \epsilon^{S_\perp p' pn} \right]
\left( 
{\hat{s}\ \widehat{\sigma}_F(\hat{s},\hat{t},\hat{u})\over \hat{t}\hat{u}} 
-{\widehat{\sigma}_I (\hat{s},\hat{t},\hat{u}) \over \hat{u}}
\right) ,
\label{SGPformula}
\end{eqnarray}
\begin{wrapfigure}{r}{0.5\columnwidth}
\vspace*{-\intextsep}
\centerline{\includegraphics[height=.15\textheight,clip]{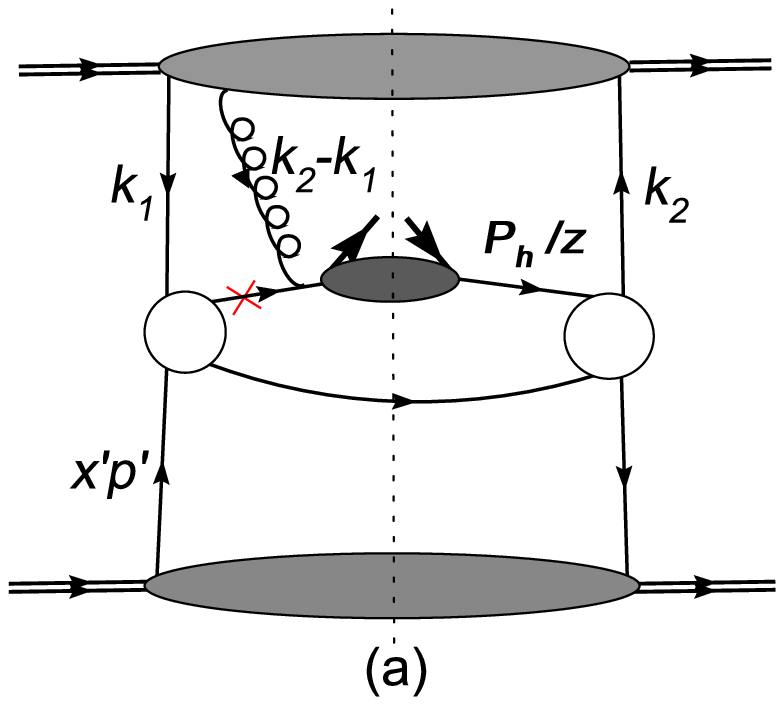}\hspace{0.05cm}
\includegraphics[height=.15\textheight,clip]{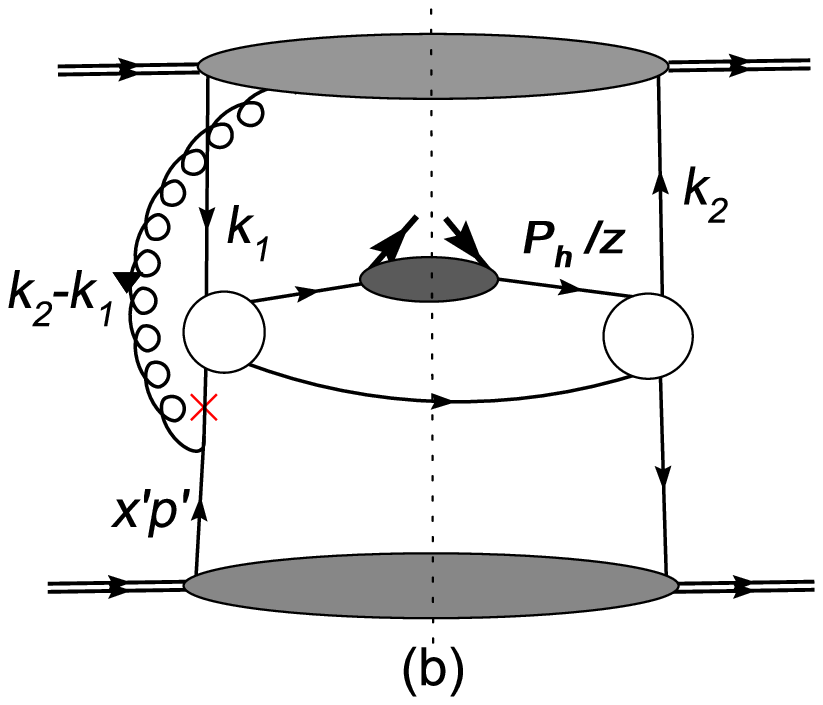}}
\caption{(a) and (b) as FSI and ISI diagrams for
SGP mechanism, respectively.
White circles denote hard scattering between quarks.}
\end{wrapfigure}
where the hard cross sections from the FSI and ISI diagrams, 
$\widehat{\sigma}_W =A_{W,1} (\hat{s}^2 +\hat{u}^2)/\hat{t}^2 +A_{W,2}
(\hat{s}^2 +\hat{t}^2)/\hat{u}^2 + A_{W,3}\hat{s}^2/(\hat{t}\hat{u})$ for $W= F$ and $I$,
are the same as the above $\widehat{\sigma}_U$, except for the associated color factors
$A_{W,i}$ that come from the {\em color insertion} of $t^a$, similarly as ${\cal B}_j$ of 
(\ref{tw3formula}).
Note that the combination, $x dG_F^q (x, x)/dx - G_F^q (x, x)$, in (\ref{SGPformula})
is a consequence of the scale invariance, 
$\widehat{\sigma}_{U}( \hat{s},\hat{t},\hat{u})
=\widehat{\sigma}_{U}(\lambda  \hat{s}, \lambda \hat{t}, \lambda \hat{u})$, similarly as in
$p^\uparrow p\to \gamma X$ discussed below (\ref{dyf}).
These remarkable structures arise universally 
for all the other relevant channels associated with the ``primordial'' 
2-to-2 partonic subprocesses, 
$q\bar{q}\to q\bar{q}$, $q\bar{q}\to gg$, 
$qg \to qg$, etc. (see also~\cite{KQVY06}).

We have derived the novel master formula which gives the twist-3 SGP contributions to the SSA
entirely in terms of the knowledge of the ``primordial'' twist-2 cross section.
This is based on the new approach, which allows us to disentangle ISI as well as FSI with
the soft coherent-gluon, 
irrespective of the details of the corresponding partonic subprocess.
Thus our single master formula is applicable universally to all processes relevant to 
SSA, 
including QED-induced processes like DY process, direct $\gamma$ production, SIDIS, etc.,
and also QCD-induced processes like $p^\uparrow p\to \pi X$, $p^\uparrow p\to 2{\rm jets}\ X$, 
$pp\to\Lambda^\uparrow X$, etc. For SSA associated with the chiral-even twist-3 function
$G_F^q(x, x)$, the primordial
twist-2 process is unpolarized as discussed above, while for SSA associated with the chiral-odd
functions, the primordial process is the polarized one~\cite{KT2}.
The primordial twist-2 structure behind the SGP mechanism manifests gauge invariance of the results, and
unveil the remarkable role of 
scale invariance.


\begin{footnotesize}




\begin{thebibliography}{99}
\bibitem{url} Slides: \\ 
\verb$http://indico.cern.ch/contributionDisplay.py?contribId=160&sessionId=4&confId=9499$
\bibitem{HC} See the articles on the data at RHIC, and those from HERMES, COMPASS,
etc. in these proceedings.
\bibitem{boer}
D.~Boer,
  Phys. Rev. {\bf D60} 014012 (1999).
\bibitem{ekt06}
  H.~Eguchi, Y.~Koike and K.~Tanaka,
Nucl. Phys. {\bf B752} 1 (2006); ibid. {\bf B763} 198 (2007).
\bibitem{KT} Y.~Koike and K.~Tanaka,
  Phys. Lett. {\bf B646} 232 (2007).
\bibitem{KT2}
  Y.~Koike and K.~Tanaka,
  arXiv:hep-ph/0703169, Phys. Rev. {\bf D} in press (2007).
\bibitem{JQVY06} 
X.D.~Ji, J.W.~Qiu, W.~Vogelsang and F.~Yuan,
Phys. Rev. {\bf D73} 094017 (2006).
\bibitem{KQVY06} 
C. Kouvaris, J.W.~Qiu, W.~Vogelsang and F.~Yuan, 
Phys. Rev. {\bf D74} 114013 (2006).
\end{thebibliography}
%

\end{footnotesize}


\end{document}